# Observational Evidence for a Correlation between the Magnetic Field of Jets and Star Formation Rate in Host Galaxies

Yongyun Chen (陈永云)[1] , Qiusheng Gu (顾秋生)[2], Junhui Fan (樊军辉)[3], Xiaotong Guo (郭晓通)[4], Xiaoling Yu (俞效龄)[1], Nan Ding (丁楠)[5], and Dingrong Xiong (熊定荣)[6]
[1] College of Physics and Electronic Engineering, Qujing Normal University, Qujing 655011, People's Republic of China; ynkmcyy@yeah,net
[2] School of Astronomy and Space Science, Nanjing University, Nanjing 210093, People's Republic of China; qsgu@nju.edu.cn
[3] Center for Astrophysics, Guangzhou University, Guangzhou 510006, People's Republic of China
[4] Anqing Normal University, Anhui 246133, People's Republic of China
[5] School of Physical Science and Technology, Kunming University, Kunming 650214, People's Republic of China
[6] Yunnan Observatories, Chinese Academy of Sciences, Kunming 650011, People's Republic of China
*Received 2024 July 09; revised 2024 November 25; accepted 2024 December 05; published 2025 January 3*

## Abstract

Accretion supermassive black holes in the center of active galaxies usually produce "jet"-collimated bipolar outflows of relativistic particles. Magnetic fields near the black hole event horizon may play a crucial role in the formation of jets/outflows. Both theory and observation indicate that jets/outflows driven by centrally active supermassive black holes have a feedback effect on the overall properties of the host galaxies. Therefore, the magnetic field is a key ingredient for the formation and evolution of galaxies. Here, we report a clear correlation between the magnetic field of jets and star formation rate for a large sample of 96 galaxies hosting supermassive black holes, which suggests that the star formation of active galactic nuclei host galaxies may be powered by the jets.

*Unified Astronomy Thesaurus concepts:* Active galactic nuclei (16); Star formation (1569); Supermassive black holes (1663); AGN host galaxies (2017)

## 1. Introduction

In the past few decades, the importance of magnetic fields in star formation has been a controversial topic. In two papers, Mestel (1965a, 1965b) argued for the importance of magnetic fields in star formation (L. Mestel 1965a, 1965b). He pointed out that the average region of the interstellar medium (ISM) containing a stellar amount of mass cannot simply collapse to stellar density because it contains too much angular momentum. He believes that the magnetic field may play a vital role in eliminating angular momentum from interstellar material to allow star formation. C. F. McKee & E. C. Ostriker (2007) suggested that magnetic fields have a major effect in extracting the angular momentum of the accreting gas and cause a large amount of accretion gas to form stars.

The magnetic field is closely related to star formation. Some authors have found that the magnetic field enhances star formation (S. Niklas & R. Beck 1997; D. J. Bomans & M. Krause 2007; K. T. Chyży 2008; K. T. Chyży et al. 2011; D. R. G. Schleicher & R. Beck 2013; V. Heesen et al. 2014; F. S. Tabatabaei et al. 2017; V. Heesen & T. L. Klocke 2023). D. R. G. Schleicher & R. Beck (2013) proposed a theoretical explanation for the observational relationship between magnetic field strength ($B$) and star formation rate (SFR) surface density ($\Sigma_{SFR}$). This $B$–SFR relation is $B \propto \Sigma_{SFR}^{1/3}$. They assumed that the kinetic energy density of a gas is regulated by star formation, so this relationship is the result of saturation of small-scale dynamo. However, as pointed out by U. P. Steinwandel et al. (2020), a similar relationship is found when the magnetic field is amplified only by adiabatic compression. Due to the rapid amplification of the magnetic field through turbulence, the small-scale dynamo is thought to have provided a strong magnetic field during the formation of the first stars and galaxies (R. Beck et al. 1996; T. G. Arshakian et al. 2009; D. R. G. Schleicher et al. 2010; J. Schober et al. 2012; M. A. Latif et al. 2013). However, F. S. Tabatabaei et al. (2018) found that there is an anticorrelation between the equipartition magnetic field and SFR for NGC 1097. They suggested the massive star formation quenching by nonthermal effects in the center of NGC 1097. At present, there has been ongoing debate on whether magnetic fields have an impact on star formation. There has always been a lack of large samples to study the relationship between magnetic fields and the SFR of active galactic nuclei (AGN) host galaxies.

Observations have found that jets/outflows have both negative (M. Cano-Diaz et al. 2012; S. Carniani et al. 2016; M. Bischetti et al. 2022; Z. Chen et al. 2022; C. Lammers et al. 2023) and positive (S. Croft et al. 2006; I. J. Feain et al. 2007; Rodriguez Zaurin et al. 2007; D. Elbaz et al. 2009; M. Imanishi et al. 2011; R. M. Crockett et al. 2012; P. C. Zinn et al. 2013; Z. Schutte & A. E. Reines 2022; G. Venturi et al. 2023) feedback effects on the host galaxies of AGN. The theoretical model of relativistic jets in AGN suggests that the jets/outflows mainly come from the spin of the black hole, the mass of the central black hole, and the magnetic field near the event horizon of the black hole (R. D. Blandford & R. L. Znajek 1977). X. Cao (2018) suggested that the jets can be accelerated by the magnetic field. Moreover, the magnetic field plays an important role in regulating the star formation process on galactic scales. The magnetic fields regulate the transport of highly energetic particles and are essential in any theory of star formation (M. R. Krumholz & C. Federrath 2019). At the same time, the magnetic field strength established by small-scale dynamo may contribute greatly to the observed correlation between SFR and magnetic field strength in galaxies

(D. R. G. Schleicher & R. Beck 2013). M. Livio et al. (1999) suggested that the large-scale field of the jet can be produced from the small-scale field created by dynamo processes. Therefore, there may be a correlation between the magnetic field of the jet and star formation. In this paper, we use a large sample of 96 galaxies hosting supermassive black holes to study the relation between the magnetic field of the jets and SFR. The second part describes the samples; the third part describes the results and discussions; and the fourth part is the conclusion. Throughout the paper, we adopt a cosmology with $H_0 = 70 \, \mathrm{km \, s^{-1} \, Mpc^{-1}}$, $\Omega_m = 0.3$, and $\Omega_\Lambda = 0.7$.

## 2. The Sample

The jets associated with these supermassive black holes have well-determined magnetic fields of the jets and the host galaxies have reliable redshifts. These galaxies hosting supermassive black holes come from the catalog of A. B. Pushkarev et al. (2012) and M. Zamaninasab et al. (2014). At the same time, we consider that these sources have multiband photometric data. We obtain the SFR and stellar mass of the host galaxies of a supermassive black hole by fitting multiband spectral energy distribution (SED) data. Table 1 lists the properties of 96 galaxies hosting supermassive black holes, including stellar mass and SFRs of host galaxies.

### 2.1. Calculating the Magnetic Field from the Core-shift Measure

According to the standard model of relativistic jets (R. D. Blandford 1979), the observation position of the radio core is a function of the observation frequency. This frequency-dependent core position shift can be used to estimate magnetic field strength and electron number density (A. P. Lobanov 1998; K. Hirotani 2005). This method is based on several simple and reasonable assumptions: the flow is conical, the half opening angle $\theta_j$ is small, and $\Gamma$ remains constant (without significant acceleration or deceleration). In addition, it is assumed that the magnetic field strength and electron number density decrease with the increase of the distance from the central engine and follow the power function relationship: $B(z) = B_{1 \, \mathrm{pc}} (z/1 \, \mathrm{pc})^{-\alpha_b}$, $n_e(z) = n_{1 \, \mathrm{pc}} (z/1 \, \mathrm{pc})^{-\alpha_n}$, where $z$ is the distance from the black hole (M. Zamaninasab et al. 2014). Here, $B_{1 \, \mathrm{pc}}$ and $n_{1 \, \mathrm{pc}}$ represent the magnetic field strength and electron number density of the jet frame at a distance of 1 pc away from the apex of the jet. These assumptions lead to the power-law behavior of the core shift as a function of the observation frequency (A. P. Lobanov 1998), $r_{\mathrm{core}} \propto \nu^{-1/k_r}$, where $r_{\mathrm{core}}$ is the absolute position of the radio core and $k_r$ is the core-shift index. The so-called "core-shift" effect can be performed via high-angular-resolution radio observations (A. P. Lobanov 1998), i.e., it shifts upstream at higher frequencies and downstream at lower frequencies (A. B. Pushkarev et al. 2012). The observed shift (in milliarcseconds) of the core between $\nu_1$ and $\nu_2$ ($\Delta r_{\mathrm{core},\nu_1\nu_2}$) is defined using $|\mathrm{CS}| = \Delta r_{\mathrm{core},\nu_1\nu_2}$, where $|\mathrm{CS}|$ is the absolute core shift. The CS is $\mathrm{CS} = \mathrm{IS} - \mathrm{OS}$ (A. B. Pushkarev et al. 2012), where IS is the image shift and OS is the difference in the coordinates (offset shift relative to the map center). If the apparent core positions measured at two frequencies $\nu_1$, $\nu_2$ ($\nu_1 < \nu_2$) differ by $\Delta r_{\mathrm{core},\nu_1\nu_2}$, the core position shift ($\Omega_{r\nu}$) is estimated using the following formula,

$$\Omega_{r\nu} = 4.85 \times 10^{-9} \left[ \frac{\Delta r_{\mathrm{core},\nu_1\nu_2} D_{\mathrm{L}}}{(1+z_*)^2} \right] \left[ \frac{\nu_1^{1/k_r} \nu_2^{1/k_r}}{\nu_2^{1/k_r} - \nu_1^{1/k_r}} \right] \mathrm{pc \, GHz}, \quad (1)$$

where $D_L$ is the luminosity distance, and $z_*$ is the redshift of sources (note we use $z_*$ for redshift to distinguish it from the cylindrical coordinate $z$).

If the magnetic field decays at a distance of $z^{-1}$ as expected by the azimuthal dominant magnetic field, and the energy ratio of magnetic particles to nonthermal particles remains constant, then it is expected that $k_r = 1$. Many observations have indeed revealed the $k_r = 1$ behavior of core movement, supporting the hypothesis that there exists an azimuthal dominant magnetic field in the radio core region (Y. Y. Kovalev et al. 2008; S. P. O'Sullivan & D. C. Gabuzda 2009; K. V. Sokolovsky et al. 2011). Assuming equipartition between magnetic energy and particle energy at the radio core position, based on the above assumption of the scaling of magnetic field and relativistic electron density with distance, the jet frame magnetic field strength at 1 pc from the black hole can be estimated through the following formula (K. Hirotani 2005; S. P. O'Sullivan & D. C. Gabuzda 2009),

$$B_{1 \, \mathrm{pc}} \simeq 0.025 \left( \frac{\Omega_{r\nu}^3 (1+z)^2}{\theta_j \delta^2 \sin^2 \theta} \right)^{1/4}, \quad (2)$$

where $\theta$ is the viewing angle, and $\delta$ is the Doppler factor. It is assumed that the jet spectral index $\alpha_{\mathrm{jet}} = -0.5$. The Doppler factor is $\delta = \Gamma^{-1}(1 - \beta \cos\theta)^{-1}$ and the apparent jet speed is $\beta_{\mathrm{app}} = \beta \sin\theta (1 - \beta \cos\theta)^{-1}$ (A. B. Pushkarev et al. 2012). For the estimate of the magnetic field, we assume that the jet is viewed at the critical angle (A. B. Pushkarev et al. 2012) $\theta \simeq \Gamma^{-1}$. This yields (A. B. Pushkarev et al. 2012)

$$B_{1 \, \mathrm{pc}} \simeq 0.042 \Omega_{r\nu}^{3/4} (1+z)^{1/2} (1+\beta_{\mathrm{app}}^2)^{1/8}. \quad (3)$$

### 2.2. The Stellar Mass and SFR of AGN Host Galaxies

We derive the stellar masses and SFRs of the host galaxies using version 2022.0 of the SED-fitting code CIGALE[7] (D. Burgarella et al. 2005; S. Noll et al. 2009; M. Boquien et al. 2019). We use the optical photometry data from Pan-STARRS1–DR1 Surveys (K. C. Chambers et al. 2016). The IR data was taken with the Two Micron All Sky Survey (2MASS; M. F. Skrutskie et al. 2006) and Wide-field Infrared Survey Explorer (WISE; E. L. Wright et al. 2010). Consequently, we have 12 photometric data ($g$, $r$, $i$, $z$, $y_{P1}$, $J$, $H$, $K/K_s$, and 3.4, 4.6, 12, and 22 $\mu$m) for the SED fitting. In our work, we use the templates of galaxy + AGN to fit the SEDs of our sample. The template for galaxies is composed of four modules, including star formation history, single stellar population model (G. Bruzual & S. Charlot 2003), dust attenuation (D. Calzetti et al. 2000), and dust emission (B. T. Draine & A. Li 2007; D. A. Dale et al. 2014). The module used for the component of AGN is J. Fritz et al. (2006). These modules are all included in CIGALE. We use the parameters and values given in Table 2 to define the CIGALE grid of model galaxy SEDs. CIGALE

---

[7] https://cigale.lam.fr/

**Table 1**
The Sample of Galaxies Hosting Supermassive Black Holes

| Name | R.A. | Decl. | $z_*$ | $B_{1\,pc}$ | log SFR$_{sed}$ | log $M_*$ | $\beta_{app}(c)$ | $\Omega_{r\nu}$ (pc GHz) |
|---|---|---|---|---|---|---|---|---|
| (1) | (2) | (3) | (4) | (5) | (6) | (7) | (8) | (9) |
| QSO B0003−066 | 1.557886944 | −6.393148611 | 0.347 | <0.2 | 2.495 | 11.26 | 2.89 | <4.55 |
| QSO B0106+013 | 17.16154611 | 1.583423056 | 2.099 | <0.79 | 3.057 | 12.58 | 26.5 | <7.86 |
| QSO B0119+115 | 20.4233125 | 11.83067 | 0.57 | 1.63 | 3.115 | 11.24 | 17.1 | 37.75 |
| QSO B0133+476 | 24.244145 | 47.85808333 | 0.859 | 0.77 | 3.381 | 11.84 | 12.98 | 13.69 |
| QSO B0149+218 | 28.07524583 | 22.11880556 | 1.32 | 1.69 | 3.594 | 11.8 | 18.55 | 29.64 |
| QSO B0202+149 | 31.21005806 | 15.23640111 | 0.405 | 0.48 | 1.999 | 10.66 | 6.41 | 10.87 |
| QSO B0212+735 | 34.37838944 | 73.82572806 | 2.367 | 1.27 | 3.389 | 12.54 | 7.64 | 21.24 |
| QSO B0224+671 | 37.20854806 | 67.35084139 | 0.523 | 0.75 | 2.605 | 11.1 | 11.63 | 15.47 |
| QSO B0238−084 | 40.269994 | −8.255764 | 0.005 | 0.015 | −0.379 | 9.01 | 0.23 | 0.226 |
| QSO B0333+321 | 54.12544833 | 32.30815056 | 1.259 | 1.95 | 3.648 | 12.11 | 12.76 | 41.51 |
| QSO B0336−019 | 54.8789075 | −1.776612222 | 0.852 | 0.92 | 3.326 | 11.51 | 22.36 | 14.41 |
| QSO B0403−132 | 61.39169917 | −13.13715694 | 0.571 | 1.54 | 2.871 | 10.46 | 19.69 | 33.3 |
| QSO B0420−014 | 65.81583639 | −1.342518056 | 0.914 | 1.41 | 3.174 | 12.61 | 7.36 | 35.94 |
| QSO B0528+134 | 82.73506972 | 13.53198583 | 2.07 | 1.6 | 3.851 | 12.04 | 19.2 | 22.71 |
| QSO B0607−157 | 92.42062306 | −15.71129778 | 0.324 | 0.68 | 2.186 | 11.16 | 3.93 | 21.22 |
| QSO B0716+714 | 110.4727019 | 71.34343417 | 0.31 | 0.49 | 3.237 | 12.68 | 10.06 | 10.16 |
| QSO B0736+017 | 114.8251414 | 1.617949444 | 0.191 | <0.2 | 1.427 | 10.92 | 14.32 | <2.94 |
| QSO B0738+313 | 115.2945969 | 31.20006361 | 0.631 | 0.81 | 3.542 | 10.55 | 10.76 | 16.85 |
| QSO B0748+126 | 117.7168572 | 12.51800806 | 0.889 | 0.84 | 3.472 | 11.64 | 18.37 | 13.49 |
| QSO B0754+100 | 119.2776789 | 9.943014444 | 0.266 | 0.88 | 2.423 | 10.51 | 14.4 | 20.38 |
| QSO B0805−077 | 122.0647333 | −7.852746111 | 1.837 | 2.71 | 4.446 | 12.02 | 50.6 | 34.84 |
| QSO B0804+499 | 122.1652764 | 49.84348083 | 1.436 | 0.48 | 1.972 | 11.45 | 1.83 | 11.5 |
| QSO B0823+033 | 126.4597431 | 3.156811389 | 0.506 | 0.83 | 3.604 | 10.77 | 17.8 | 15.57 |
| QSO B0827+243 | 127.7170258 | 24.18328333 | 0.94 | 1.18 | 2.981 | 12 | 22.01 | 19.64 |
| QSO B0836+710 | 130.3515219 | 70.89504806 | 2.218 | 1.93 | 3.349 | 12.27 | 25.38 | 25.72 |
| QSO B0851+202 | 133.7036456 | 20.10851139 | 0.306 | <0.28 | 2.989 | 11.7 | 15.17 | <4.18 |
| QSO B0859−140 | 135.5701458 | −14.25859472 | 1.339 | 1.7 | 3.922 | 10.31 | 16.47 | 30.96 |
| QSO B0906+015 | 137.2920483 | 1.359893889 | 1.024 | 1.61 | 3.522 | 12.59 | 20.68 | 29.45 |
| QSO B0917+624 | 140.4009625 | 62.26449444 | 1.446 | 1.09 | 3.245 | 10.79 | 15.57 | 16.99 |
| QSO B0923+392 | 141.7625581 | 39.03912556 | 0.695 | <0.33 | 2.525 | 12.13 | 4.29 | <6.64 |
| QSO B0945+408 | 147.2305756 | 40.66238528 | 1.249 | 1.1 | 3.789 | 12.01 | 18.6 | 17.02 |
| QSO B0953+254 | 149.2078142 | 25.25445833 | 0.712 | <0.42 | 2.854 | 10.89 | 11.52 | <6.71 |
| QSO B1015+359 | 154.5457458 | 35.71096139 | 1.226 | 0.92 | 3.689 | 10.95 | 12.46 | 15.55 |
| IVS B1036+054 | 159.6949164 | 5.208079444 | 0.473 | 0.74 | 1.609 | 11.47 | 6.15 | 19.26 |
| QSO B1038+064 | 160.3215106 | 6.171367778 | 1.265 | 1.19 | 3.501 | 9.92 | 11.87 | 21.97 |
| QSO B1045−188 | 162.0275858 | −19.15992417 | 0.595 | 0.86 | 3.087 | 11.11 | 8.57 | 19.94 |
| QSO B1101+384 | 166.1138081 | 38.20858306 | 0.031 | 0.1 | 1.291 | 8.95 | 0.82 | 2.81 |
| QSO B1127−145 | 172.5293639 | −14.82428444 | 1.184 | 0.84 | 2.984 | 12.5 | 14.18 | 13.24 |
| QSO B1150+812 | 178.30208 | 80.974765 | 1.25 | 0.68 | 2.608 | 11.25 | 7.09 | 12.34 |
| QSO B1156+295 | 179.88264 | 29.24551 | 0.73 | 1.17 | 3.367 | 12.81 | 24.73 | 20.11 |
| QSO B1219+285 | 185.38204 | 28.23292 | 0.102 | 0.08 | −0.956 | 11.38 | 4.05 | 1.39 |
| QSO B1219+044 | 185.5939317 | 4.221062222 | 0.965 | 0.81 | 2.725 | 10.96 | 2.35 | 24.16 |
| QSO B1222+216 | 186.2268733 | 21.37957778 | 0.432 | 0.9 | 3.296 | 11.06 | 21.1 | 17.03 |
| QSO B1228+126 | 187.70593 | 12.391123 | 0.004 | 0.012 | −1.662 | 8.86 | 0.24 | 0.132 |
| QSO B1253−055 | 194.0465275 | −5.7893125 | 0.536 | <0.42 | 2.754 | 12.93 | 20.57 | <5.88 |
| QSO B1302−102 | 196.3875625 | −10.55539639 | 0.278 | 0.7 | 2.858 | 10.72 | 5.41 | 20.34 |
| QSO B1308+326 | 197.6194328 | 32.34549528 | 0.997 | 0.96 | 2.981 | 11.85 | 27.17 | 13.61 |
| QSO B1334−127 | 204.4157614 | −12.95685917 | 0.539 | 1.23 | 3.496 | 11.52 | 10.26 | 31.08 |
| QSO B1413+135 | 213.9950731 | 13.33992028 | 0.247 | 0.44 | 2.539 | 10.45 | 1.8 | 15.7 |
| QSO B1458+718 | 224.7815733 | 71.67218139 | 0.904 | 0.51 | 2.298 | 11.53 | 7.04 | 9.46 |
| QSO B1502+106 | 226.1040825 | 10.49422194 | 1.839 | 0.69 | 3.345 | 12.08 | 14.77 | 8.5 |
| QSO B1504−166 | 226.7700294 | −16.87524111 | 0.876 | 0.66 | 1.865 | 11.38 | 4.31 | 15.9 |
| QSO B1508−055 | 227.7232975 | −5.718726944 | 1.191 | 1.52 | 3.449 | 11.69 | 18.64 | 26.81 |
| QSO B1510−089 | 228.2105539 | −9.099952778 | 0.36 | 0.73 | 2.755 | 12.8 | 20.14 | 13.5 |
| QSO B1532+016 | 233.7185569 | 1.517835278 | 1.42 | 1.14 | 3.262 | 11.98 | 14.11 | 18.81 |
| QSO B1538+149 | 235.2062139 | 14.79609 | 0.605 | 0.49 | 2.904 | 11.27 | 8.73 | 9.25 |
| QSO B1546+027 | 237.3726533 | 2.61699 | 0.414 | <0.32 | 2.814 | 11.4 | 12.08 | <5.09 |
| QSO B1606+106 | 242.1925133 | 10.48549333 | 1.226 | 0.79 | 2.817 | 12.46 | 18.91 | 10.97 |
| QSO B1633+382 | 248.8145542 | 38.13458361 | 1.814 | 1.62 | 3.702 | 13.24 | 29.45 | 21.21 |
| QSO B1637+574 | 249.5560681 | 57.33999417 | 0.751 | 0.71 | 3.308 | 11.97 | 10.61 | 13.51 |
| QSO B1642+690 | 250.5327022 | 68.94437667 | 0.751 | <0.48 | 1.785 | 11.32 | 16.65 | <6.85 |
| QSO B1641+399 | 250.7450414 | 39.81027611 | 0.593 | 1.2 | 2.968 | 11.57 | 19.27 | 23.85 |

**Table 1**
(Continued)

| Name | R.A. | Decl. | $z_*$ | $B_{1\,pc}$ | log SFR$_{sed}$ | log $M_*$ | $\beta_{app}(c)$ | $\Omega_{r\nu}$ (pc GHz) |
|---|---|---|---|---|---|---|---|---|
| (1) | (2) | (3) | (4) | (5) | (6) | (7) | (8) | (9) |
| QSO B1652+398 | 253.4675694 | 39.76016917 | 0.033 | 0.1 | −0.377 | 10.97 | 0.21 | 3.25 |
| QSO B1655+077 | 254.5375478 | 7.690983611 | 0.621 | 0.47 | 1.174 | 10.82 | 14.45 | 7.45 |
| QSO B1726+455 | 261.8652117 | 45.51103667 | 0.717 | <0.28 | 0.603 | 10.86 | 1.82 | <6.73 |
| QSO B1730-130 | 263.2612739 | −13.08043 | 0.902 | 1.69 | 3.166 | 9.35 | 35.69 | 27.31 |
| QSO B1749+701 | 267.136835 | 70.09743556 | 0.77 | 1.04 | 3.017 | 12.32 | 6.03 | 27.03 |
| QSO B1749+096 | 267.8867439 | 9.6502025 | 0.322 | 0.33 | 2.207 | 12.27 | 6.84 | 6.92 |
| QSO B1803+784 | 270.19035 | 78.46778306 | 0.68 | <0.39 | 2.738 | 12.18 | 8.97 | <6.58 |
| QSO B1807+698 | 271.7111694 | 69.82447472 | 0.051 | 0.12 | 0.656 | 10.34 | 0.1 | 3.81 |
| QSO B1823+568 | 276.0294503 | 56.850415 | 0.664 | 0.74 | 2.804 | 11.02 | 20.85 | 11.79 |
| QSO B1828+487 | 277.3821867 | 48.74637556 | 0.692 | 0.69 | 2.431 | 12.31 | 13.65 | 12.24 |
| QSO B1845+797 | 280.53746 | 79.77142 | 0.057 | 0.089 | 1.2 | 10.35 | 2.29 | 1.65 |
| QSO B1849+670 | 282.3169681 | 67.09491111 | 0.657 | 0.52 | 2.422 | 11.71 | 30.63 | 6.48 |
| QSO B1928+738 | 291.9520633 | 73.96710278 | 0.302 | 0.54 | 2.837 | 10.95 | 8.43 | 12.31 |
| QSO B2005+403 | 301.9372708 | 40.49683639 | 1.736 | 2.34 | 3.815 | 12.15 | 12.21 | 47.18 |
| QSO B2037+511 | 309.6543114 | 51.32018417 | 1.686 | <0.45 | 1.534 | 12.52 | 3.3 | <7.98 |
| QSO B2113+293 | 318.8725564 | 29.5606575 | 1.514 | 1.16 | 4.024 | 11.6 | 1.4 | 37.78 |
| QSO B2121+053 | 320.9354892 | 5.589470278 | 1.941 | 1.43 | 3.776 | 12.71 | 13.29 | 22.61 |
| QSO B2128-123 | 322.8969239 | −12.11799889 | 0.501 | 0.98 | 3.35 | 11.84 | 6.94 | 26.41 |
| QSO B2131-021 | 323.5429564 | −1.888121944 | 1.285 | 1.02 | 3.081 | 11.41 | 20.02 | 14.94 |
| QSO B2134+004 | 324.1607756 | 0.698394444 | 1.932 | 1.31 | 3.463 | 12.25 | 5.94 | 26.23 |
| QSO B2136+141 | 324.7554553 | 14.39333111 | 2.427 | <0.55 | 2.871 | 12.41 | 5.43 | <7.68 |
| QSO B2145+067 | 327.0227444 | 6.960723333 | 0.99 | <0.34 | 2.91 | 12.16 | 2.52 | <7.48 |
| QSO B2155-152 | 329.5261744 | −15.01925778 | 0.672 | 1.89 | 3.386 | 11.97 | 18.11 | 43.23 |
| QSO B2200+420 | 330.6803808 | 42.27777222 | 0.069 | 0.09 | 1.785 | 10.83 | 10.57 | 1.21 |
| QSO B2201+315 | 330.8123986 | 31.76063194 | 0.295 | 0.95 | 2.51 | 10.7 | 7.87 | 27.04 |
| QSO B2201+171 | 330.8620569 | 17.43006889 | 1.076 | 1.55 | 3.775 | 13.18 | 2.55 | 54.08 |
| QSO B2209+236 | 333.0248597 | 23.92792889 | 1.125 | 0.39 | 1.357 | 11.63 | 3.43 | 7.69 |
| QSO B2223-052 | 336.4469136 | −4.950386389 | 1.404 | 1.47 | 3.005 | 13.03 | 17.34 | 24.63 |
| QSO B2227-088 | 337.4170181 | −8.548454444 | 1.56 | 1.44 | 3.76 | 12.6 | 8.14 | 29.6 |
| QSO B2230+114 | 338.15173 | 11.73076917 | 1.037 | 2.12 | 4.477 | 12.25 | 15.41 | 46.48 |
| QSO B2243-123 | 341.5759875 | −12.11424778 | 0.632 | 0.78 | 2.049 | 11.17 | 5.49 | 20.1 |
| QSO B2251+158 | 343.4906167 | 16.14821139 | 0.859 | 1.13 | 3.536 | 12.17 | 14.19 | 22 |
| QSO B2345-167 | 357.0108689 | −16.52000611 | 0.576 | 0.9 | 2.736 | 10.29 | 13.45 | 18.44 |
| QSO B2351+456 | 358.5903392 | 45.88450778 | 1.986 | 1.72 | 3.729 | 11.67 | 18.01 | 25.97 |

**Note.** Column (1): name; column (2): the R.A. in decimal degrees; column (3): (delineation) in decimal degrees; column (4): redshift; column (5): magnetic field strength 1 pc away from the jet base (in units Gauss); column (6): logarithm of star formation rate is estimated by fitting SED; column (7): logarithm of stellar mass; column (8): apparent velocity of the jet; column (9): core-shift measure.

identifies the best-fit SED model by minimizing the $\chi^2$ statistic. We present the example of the best SED fitting in Figure 1.

### 3. Results and Discussions

We use linear regression to analyze the relationship between the SFR and jet magnetic field in galaxies hosting supermassive black holes. Figure 2 shows the relation between SFR and jet magnetic field in our sample galaxies. A fit to the relation between the two variables is done using linmix[8] (B. C. Kelly 2007). We assume that there is a significant correlation between the two variables when $p < 0.01$. There is a significant correlation between SFR and the magnetic field of the jet for our sample galaxies (Pearson correlation coefficient $r = 0.87$, $p = 5.01 \times 10^{-30}$),

$$\log B_{1\,pc} = 0.33(\pm 0.02) \log SFR_{sed} - 1.08(\pm 0.05). \quad (4)$$

The tests of Spearman ($r = 0.82$, $p = 1.64 \times 10^{-24}$) and Kendall tau ($r = 0.62$, $p = 3.19 \times 10^{-19}$) also show a significant correlation between SFR and magnetic field of the jets for our sample galaxies. Our results suggest that the star formation may occur along the trajectory of the jet. Theoretical models indicate that relativistic jets rely on the magnetic field of black holes (e.g., R. D. Blandford & R. L. Znajek 1977). Our results may imply that jets induce star formation. P. C. Zinn et al. (2013) found that AGN with pronounced radio jets exhibit a much higher SFR, which implies that positive AGN feedback plays an important role (jet-induced star formation). V. Gaibler et al. (2012) found jet-induced star formation through numerical simulation of feedback from AGN.

The relation between SFR and stellar mass for our sample galaxies is shown in Figure 3. The yellow and green lines show the main sequences of star formation at $z \sim 1$ and $z \sim 0$ (D. Elbaz et al. 2007), respectively. From Figure 3, we find that almost all sources are located above the main sequence of star formation. This indicates that the host galaxies of these sources have high SFRs and may be undergoing star formation. We also estimate the specific SFR (sSFR = SFR/$M_*$). The relation between the sSFR and magnetic field of the jets for our sample galaxies is shown in Figure 4. We also use linear regression to analyze the relationship between the sSFR and jet magnetic field in galaxies hosting supermassive black holes. There is a significant correlation between sSFR and the magnetic field of the jet for our sample galaxies (Pearson correlation coefficient $r = 0.47$, $p = 1.18 \times 10^{-6}$),

$$\log B_{1\,pc} = 0.19(\pm 0.04)\log sSFR_{sed} + 1.52(\pm 0.33). \quad (5)$$

---

[8] https://linmix.readthedocs.io/en/latest/

Table 2
CIGALE Grid Parameter Values Adopted for the Modeling Described

| Parameter | Values | Description |
|---|---|---|
| | Star formation history: Delayed | |
| $\tau_{main}$ | 5–8000 (in steps of 10) | e-folding time of the main stellar population model (Myr). |
| Age | 200–13,260 (in steps of 10) | Age of the oldest stars in the galaxy (Myr). |
| | Single-age stellar population (SSP): G. Bruzual & S. Charlot (2003) | |
| imf | 1 | Initial mass function (G. Chabrier 2003) |
| Metallicity | 0.02 | Solar |
| Separation Age | 10 | Age of the separation (to differentiate) between the young and the old star populations (Myr). |
| | Dust attenuation: D. Calzetti et al. (2000) | |
| $E(B-V)_{young}$ | 0.001, 0.01, 0.015, 0.02, 0.04, 0.06, 0.1, 0.2, 0.3, 0.4, 0.5, 0.6, 0.7, 0.8, 0.9, 1.0 | Color excess of the stellar continuum light for the young population. |
| | Dust emission: D. A. Dale et al. (2014) | |
| $\alpha$ | 2.0, 2.1, 2.5 | Alpha from the power-law distribution ($dM_d \propto U^{-\alpha} dU$, with $M_d$ being the dust mass, and $U$ the radiation field intensity). |
| | AGN model: J. Fritz et al. (2006) | |
| $R_{max}/R_{min}$ | 10.0, 30.0, 60.0, 100.0, 150.0 | Ratio of the maximum to minimum radii of the dust torus. |
| $\tau$ | 0.1, 0.3, 0.6, 1.0, 2.0, 3.0, 6.0, 10.0 | Optical depth at 9.7 $\mu$m. |
| $\beta$ | −1.00, −0.75, −0.50, −0.25, 0.00 | Beta from the power-law density distribution for the radial component of the dust torus (Equation (3) of J. Fritz et al. 2006). |
| $\gamma$ | 0.0, 2.0, 4.0, 6.0 | Gamma from the power-law density distribution for the polar component of the dust torus (Equation (3) of J. Fritz et al. 2006). |
| Opening Angle ($\theta$) | 60.0, 100.0, 140.0 | Full opening angle of the dust torus (Figure 1 of J. Fritz et al. 2006). |
| $\psi$ | 0.001, 10.1, 20.1, 30.1, 40.1, 50.1, 60.1, 70.1, 80.1, 89.990 | Angle between equatorial axis and line of sight. |
| $f_{AGN}$ | 0.0, 0.05, 0.1, 0.15, 0.2, 0.25, 0.3, 0.35, 0.4, 0.45, 0.5, 0.55, 0.6, 0.65, 0.7, 0.75, 0.8, 0.85, 0.9, 0.95, 0.99 | Fraction of AGN torus contribution to the IR luminosity (fracAGN in Equation (1) of L. Ciesla et al. 2015) |

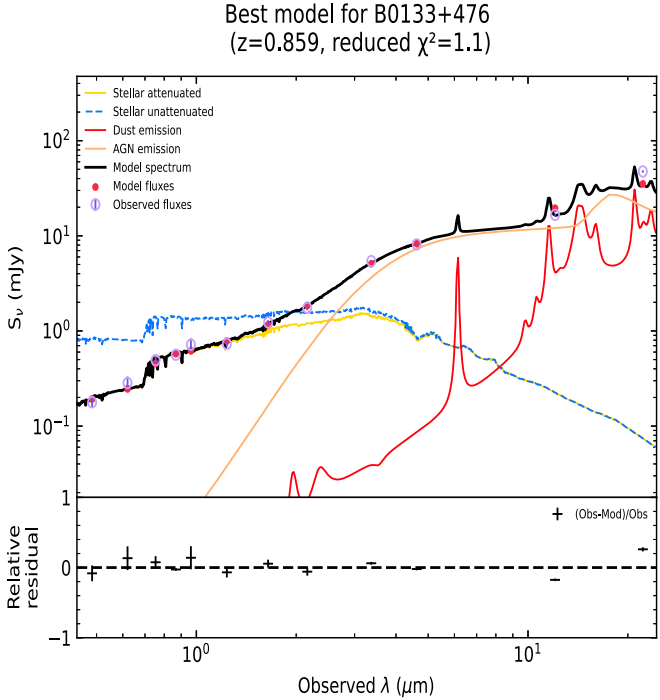

Figure 1. The example of broadband SEDs of AGN is modeled by using CIGALE.

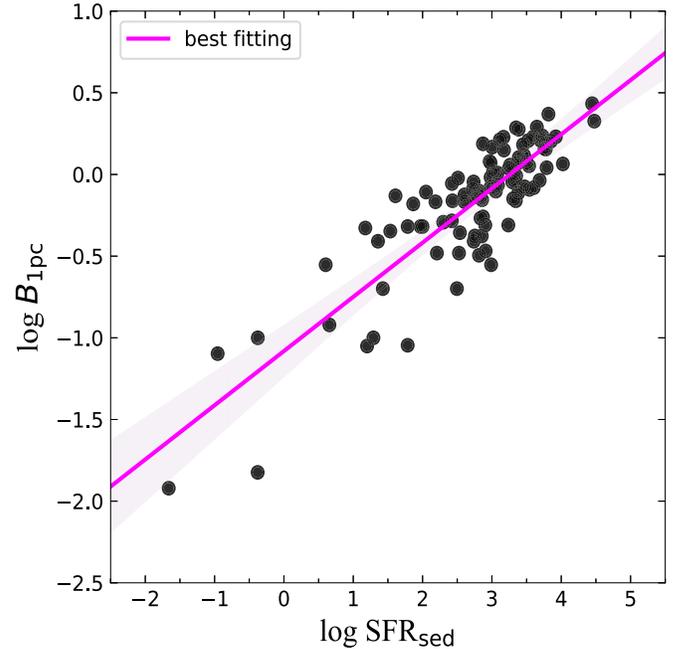

Figure 2. Relation between star formation rate and the magnetic field of the jets for galaxies hosting supermassive black holes. The star formation rate is estimated by fitting SED using CIGALE. The $B_{1\,pc}$ is the jet's comoving frame magnetic field measured by the core-shift effect. Shaded red colored areas correspond to $3\sigma$ confidence bands. The red line is the best fit.

The tests of Spearman ($r = 0.46$, $p = 1.92 \times 10^{-6}$) and Kendall tau ($r = 0.32$, $p = 4.63 \times 10^{-6}$) also show a significant correlation between sSFR and magnetic field of the jets for our sample galaxies. Our results may provide a new clue for future research about the influence of magnetic fields on star formation using theory and numerical simulations.

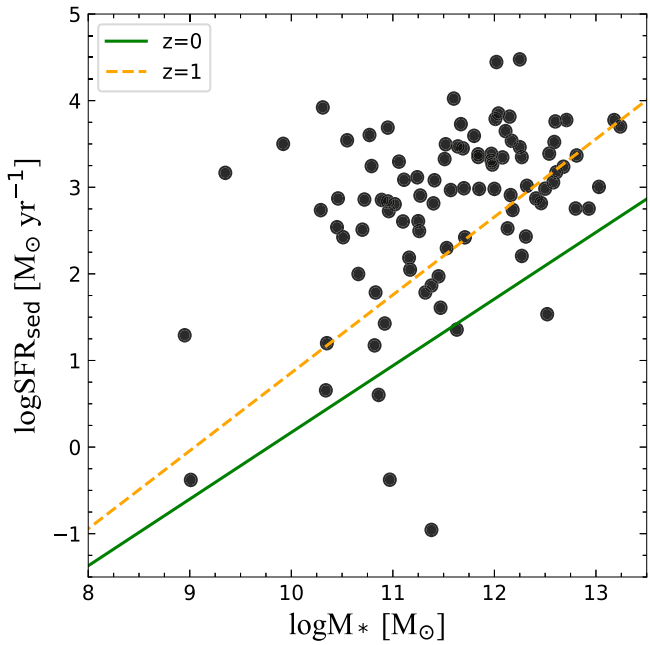

**Figure 3.** Relation between star formation rate and stellar mass for galaxies hosting supermassive black hole. The yellow and green lines show the main sequences of star formation at $z = 1$ and $z = 0$ (D. Elbaz et al. 2007), respectively.

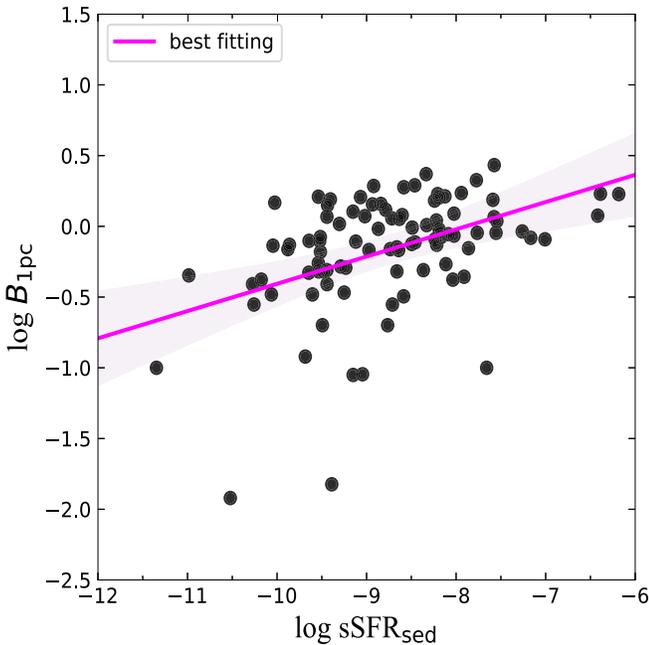

**Figure 4.** Relation between specific star formation rate and magnetic field for galaxies hosting supermassive black holes. Shaded red colored areas correspond to $3\sigma$ confidence bands. The red line is the best fit.

## 4. Conclusions

We fit multiband data to obtain a large sample of star masses and SFRs of galaxies hosting supermassive black holes and study the relationship between SFRs and the magnetic field of the jets. Our main results are as follows:

(1) We find a significantly strong correlation between the magnetic field of the jets and SFR. The magnetic field strength is proportional to the SFR, $B_{1\,pc} \propto SFR_{sed}^{0.33\pm0.02}$. Our results suggest that the jets may induce star formation of AGN host galaxies.

(2) There is also a significantly strong correlation between the magnetic field of the jets and sSFR. The magnetic field strength is proportional to the sSFR, $B_{1\,pc} \propto sSFR_{sed}^{0.19\pm0.04}$.

## Acknowledgments

Yongyun Chen is grateful for financial support from the National Natural Science Foundation of China (No. 12203028). Yongyun Chen is grateful for funding for the training Program for talents in Xingdian, Yunnan Province (2081450001). QSGU is supported by the National Natural Science Foundation of China (12121003, 12192220, and 12192222). We also acknowledge the science research grants from the China Manned Space Project with NO. CMS-CSST-2021-A05. This work is supported by the National Natural Science Foundation of China (11733001, U2031201 and 12433004). Xiaotong Guo acknowledge the support of National Nature Science Foundation of China (Nos 12303017). This work is also supported by Anhui Provincial Natural Science Foundation project number 2308085QA33. D.R.X. is supported by the NSFC 12473020, Yunnan Province Youth Top Talent Project (YNWR-QNBJ-2020-116) and the CA. Light of West China Program.

## ORCID iDs

Yongyun Chen (陈永云)  https://orcid.org/0000-0001-5895-0189